# Photoresponse in large area multi-walled carbon nanotube/ polymer nanocomposite films

Paul Stokes[1,2], Liwei Liu[1,2], Jianhua Zou[1,3,4], Lei Zhai[1,3,4], Qun Huo[1,3,4] and Saiful I. Khondaker,[1,2,*]

[1] Nanoscience Technology Center, [2] Department of Physics, [3] Department of Mechanical, Materials and Aerospace Engineering, and [4] Department of Chemistry, University of Central Florida, Orlando, Florida 32826, USA.

**Abstract**

We present a near IR photoresponse study of large area multi-walled carbon nanotube/poly(3-hexylthiophene)-block-polystyrene polymer (MWNT/P3HT-b-PS) nanocomposite films for different loading ratio of MWNT into the polymer matrix. We show that the photocurrent strongly depends on the position of the laser spot with maximum photocurrent occurring at the metal – film interface. In addition, compared to the pure MWNT film, the photoresponse is much larger in the MWNT/polymer composite films. The time constant for the photoresponse is slow and varies between 0.6 and 1.2 seconds. We explain the photoresponse by Schottky barrier modulation at the metal – film interface.

* To whom correspondence should be addressed. E-mail: saiful@mail.ucf.edu

Carbon nanotubes (CNTs) are considered to be promising building blocks for nanoelectronic and optical devices due to their special geometry, high electrical conductivity, exceptional mechanical and optical properties [1-3]. Recently, photoresponse of CNTs (both in visible and near infrared (NIR) regime) have generated considerable debate in terms of whether the photoresponse is (i) due to photon induced charge carrier (excitonic), (ii) due to heating of the CNT network (bolometric), or (iii) caused by photodesorption of oxygen molecules at the surface of the CNT. In addition, the role of the metal - CNT contact effect on the photoresponse has also been debated. In individual semiconducting single-walled carbon nanotube (SWNT) field effect transistor (FET) devices, Freitag *et al.* and Qiu *et al.* explained the photoresponse in the NIR regime using an exciton model [4,5]. Chen *et al.* showed that in individual SWNT FET device, the reduction of conductivity upon UV illumination is due to desorption of molecular oxygen from the CNT surface which causes a reduction in hole carriers [6]. In contrast, for a large area SWNT film suspended in vacuum, it was argued by Itkis *et al.* that the NIR photoresponse was due to a bolometric effect, a change in conductivity due to heating of the SWNT network [7]. In a microscopic SWNT film, Levitsky *et al* also found molecular photodesorption to be responsible for change in conductivity upon near IR illumination [8]. Other studies in macroscopic SWNT film show that the maximum photoresponse occurs at the CNT film-metal interface and that the response varies with the position of laser illumination [9-12]. This effect has been explained using exciton model where CNT-metal interface helps to separate electron and holes to induce a photovoltage. A recent study has shown that muti-walled carbon nanotube (MWNT) film also displays a positional dependent photocurrent [13]. Although there are many studies for the photoresponse in individual CNTs and CNT films, there are almost no studies on CNT/Polymer nanocomposites. Nanocomposites may provide enhanced functionality as they form the basis of new advanced materials with tailored properties. Their ease of processibility in solution, multifunctionality, flexibility, and low cost of fabrication makes them attractive candidates for large area optoelectronic devices [14-16].

In this letter, we present a near IR photoresponse study of multi-walled carbon nanotube/poly(3-hexylthiophene)-block-polystyrene (MWNT/P3HT-b-PS) polymer nanocomposite film for different loading ratio of MWNT in polymer matrix. We found that upon NIR illumination, the photocurrent either increases, remain almost zero, or decreases based on the position of the illumination with respect to electrodes with maximum photocurrent occurring when illuminated at the metal-film interface. The photoresponse [(light current − dark current)/dark current] or [$(I_{light}-I_{dark})/I_{dark}$] is up to 164% when MWNT is embedded into P3HT-b-PS while for pure MWNT film it is only 4%. The time constant of the photoresponse is rather slow, between 0.6-1.2 seconds. We explain the contact dependence photoresponse using excitonic picture and discuss reasons for slow time response. This work shows promising novel route for the fabrication of large area low cost infrared photo detectors and position sensitive detectors.

MWNT/P3HT-b-PS composites were prepared following our recently developed technique [15]. The diameter and length of the MWNTs (purchased from Nanolab) is ~10-20 nm and 5-20 μm, respectively, determined by SEM and TEM (not shown here). The appropriate amount of MWNTs and P3HT-b-PS were first dispersed into 5 ml of chloroform and sonicated in water bath kept constant around 10-15$^0$ C for 30 minutes. The resulting dispersion was very uniform and stable. Next, the appropriate amount of the solution was loaded into 10% PS in chloroform and put on a shaker for ~5 minutes for mixing. Solutions with MWNT loading ratios of 0.25, 0.50, 0.75, 1.00, 1.25, 1.50, 2.00, 3.00, 5.00, 10.00 % were prepared. In addition a pure P3HT-b-PS film and a pure MWNT film was also prepared for control experiments. The solution was then drop cast onto the glass substrate using a draw-down bar and the solvent was allowed to evaporate in a fume hood for a few hrs. Resulting films were ~40-60 μm thick. Finally, ~ 60 nm thick Au electrodes spaced $L$~3 mm and a width of $W$~20 mm were thermally evaporated using a shadow mask.

Figure 1a shows a cartoon of a final device and the electrical transport measurement setup. The room temperature dc transport measurements of the composite films was carried out using a standard two-probe technique both in dark and under illumination by a laser spot positioned at three different locations (Figure 1a). A corresponds to illumination on the left electrode/film interface, B is between the electrodes in the middle of the sample, and C is the right electrode/film interface. The near IR photo source consists of a semiconductor laser diode with peak wavelength of 806 nm (1.54 eV) driven by a Keithley 2400. The spot size of the laser was approximately 10 mm long and 1 mm wide. The photo intensity is monitored with a calibrated silicon photodiode (Thorlabs S121 B). The power intensity of the laser is ~ 5 mW/mm$^2$ at the distance it was placed from the sample (~2.5 cm). Photocurrent was measured under small $V_{bias}$ (1mV) unless mentioned otherwise. Data was collected by means of LabView interfaced with the data acquisition card and current preamplifier (DL instruments: Model 1211) capable of measuring sub pA signal.

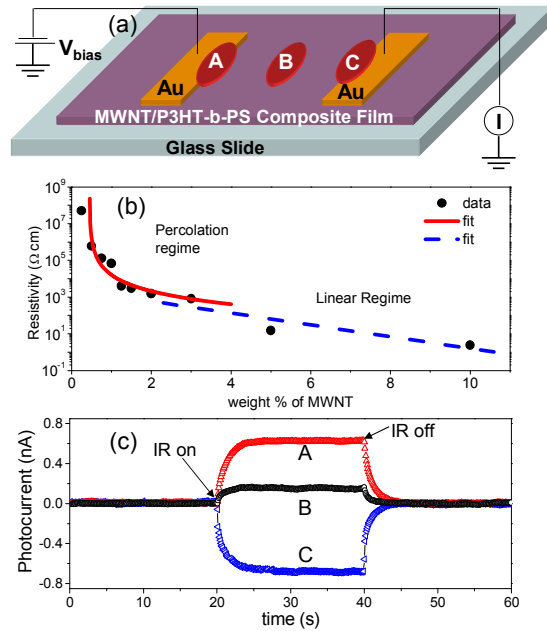

**Figure 1:** (Color online) (a) Cartoon of the device and transport measurement setup. The spacing between the electrodes is ~ 3 mm, length is ~ 20 mm, and IR laser wavelength is 806 nm. (b) Resistivity of the film versus weight percentage of MWNT in the polymer matrix. The resistivity data follows a standard percolation threshold curve $\rho \propto (p-p_c)^\alpha$ (dashed line) with $p_c$=0.45% and $\alpha$=1.95 and then reduces linearly with high weight percent. (c) Representative photocurrent, as a function of time ($t$) for 1.5% film under IR illumination at positions A, B, and C ($V_{bias}$=1mV). The IR laser is turned on at $t$=20s, left on for 20s, and then turned of at $t$=40s.

Figure 1b shows the resistivity of the sample for different loading ratio of the MWNT in P3HT-b-PS matrix measured in dark. By varying the fraction of MWNT added into the solution the electrical resistivity can be controlled by several orders of magnitude. The resistivity of the films follow the standard percolation formula $\rho \propto (p-p_c)^{-\alpha}$ where $p_c$ is the critical volume fraction and $\alpha$ is the critical exponent. From the fit to our data, we obtain $p_c = 0.45\%$ and $\alpha = 1.95$. For large concentration of MWNT, the resitivity reduces linearly. The critical exponent we obtain is within reasonable agreement for what is expected from theoretical predictions [17].

Figure 1c shows a typical photoresponse curve for one of our composite films (1.5% MWNT) where we plot photocurrent as a function of time (*t*) measured under a constant bias voltage of 1 mV when the laser spot is positioned at A, B, and C. The IR source is turned on at *t*=20 seconds, left on for 20 seconds, and then turned off at *t* =40 seconds. The current under no illumination (dark current) was subtracted from the photocurrent measurements. Three features can be noticed from this figure, (a) the photocurrent is positional dependent, (b) there is a large enhancement of photocurrent at the metal-film interface, and (c) the response time is rather slow. When illuminated at position A there is an increase in photocurrent. Position B also shows an increase in current but much smaller, whereas position C shows a decrease in photocurrent when illuminated by the IR source. Similar behaviour of the photocurrent has been observed in all our samples with different MWNT loading ratio excluding the pure

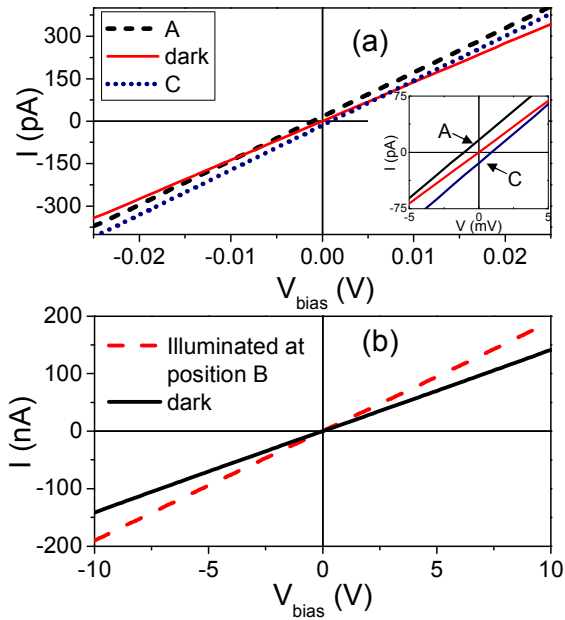

**Figure 2:** (Color online) (a) Current-voltage characteristics of a 0.5 wt% MWNT composite film illuminated at positions A, C and in the dark. Inset: Zoomed in view around the origin. (b) Full range current-voltage characteristics in the dark and when illuminated at B.

**Table 1:** Dark current, light current, photoresponse, and external quantum efficiency for MWNT/P3HT-b-PS nanocomposite films at position A. The current were measured at 1mV bias.

| wt% of MWNT | $I_{dark}$ (A) | $I_{light}$ (A) | $(I_{light}-I_{dark})/I_{dark}$ (%) | EQE |
|---|---|---|---|---|
| 0.5 | 1.27E-11 | 3.35E-11 | 164 | 6.40E-07 |
| 0.75 | 5.26E-11 | 6.72E-11 | 27.8 | 4.50E-07 |
| 1 | 1.07E-10 | 1.33E-10 | 24.3 | 8.00E-07 |
| 1.25 | 1.78E-09 | 2.10E-09 | 18.0 | 9.90E-06 |
| 1.5 | 3.20E-09 | 3.80E-09 | 18.8 | 1.80E-05 |
| 2 | 6.60E-09 | 8.20E-09 | 24.2 | 4.90E-05 |
| 3 | 1.18E-08 | 1.32E-08 | 11.9 | 4.30E-05 |
| 5 | 1.10E-06 | 1.31E-06 | 19.1 | 6.50E-03 |
| 10 | 3.14E-06 | 3.80E-06 | 21.0 | 2.00E-02 |
| 100 | 4.65E-05 | 4.84E-05 | 4.1 | 5.90E-02 |

P3HT-b-PS film and the 0.25% MWNT loading ratio. For these two samples, under 1 mV bias, signal to noise ratio is extremely low, therefore the photocurrent was not very reproducible. The S/N ratio for 0.5% composite film was 21 while for all other samples it was ~70 or higher making the detection of photocurrent easier. Table 1 summarizes experimental data from all our samples, where we show dark current, light current, photoresponse, and external quantum efficiency (EQE) for different loading ratio of MWNT composite film at position A. The EQE at $\lambda = 806$ nm at a fixed power of 5 mW was calculated using $EQE = (R_\lambda/\lambda) \cdot 1240 \ W \cdot nm/A$, where $R_\lambda$ is the responsivity defined by the photocurrent per watt of input power. As can be seen from this table, the photoresponse is 164% for 0.5% MWNT/P3HT-b-PS composite film while it is only 4.1% for pure MWNT film. The stronger photoresponse in MWNT/P3HT-b-PS films could be due to semi-conducting nature of these films compared to pure MWNT film where metallic pathways dominate. We note that in SWNT-polycarbonate composite film studied by Pradhan et al. [18] such photoresponse was also observed. However, the maximum photoresponse was only 5% and could be due to the illumination of IR around the middle part of their sample and that the effect of contact was not checked.

In order to examine the position dependent photocurrent further, we measured the current-voltage (I-V) characteristics for all of our samples. Figure 2a shows a representative I-V curve of a 0.5% MWNT composite film when illuminated at position A, position C, and in the dark. The dark I-V curve runs directly though the origin, whereas when illuminated on position A and C the I-V curve is slightly shifted above or below the origin respectively. At zero applied bias, there is about +15 pA current at A and -15 pA at C. This is more clearly seen in Figure 2a's inset. Very little or no photocurrent at zero bias was observed at position B. The zero bias photocurrent increases in magnitude at position A and C for increasing MWNT loading ratio. It is therefore clear that the finite current at zero bias must be caused by a photoinduced voltage related to metal composite film interface. It can also be seen from Figure 2a that, beyond certain critical positive applied bias the photocurrent is always positive at all positions. This means that if one applies source drain bias beyond this voltage, the contact effect may become non-noticeable. The critical voltage increases from 5 mV to 50 mV from lowest to highest weight percent in our samples. This may be a measure of the Schottky barrier in our samples. Figure 2b shows the dark and light I-V curve up to 10 V when the sample is illuminated at B. Here we do not see any offset at zero bias when illuminated.

We now examine the slow rise and decay of photocurrent when NIR source is switched on and off. Figure 3 is a plot of current as a function of time for 1.5% composite for two cases: (i) the IR is turned on at $t$ =10s, held on for 10s, and turned off at $t$=20s at position A with $V_{bias}$= 1mV, and (ii) the bias is increase from 1.0 mV to 1.2 mV at $t$=30s and then back to 1.0 mV at $t$ = 40s. When illuminated by the IR source the sample responds slowly until it reaches its steady state, and slowly recovers to the dark state current when the laser is turned off. In contrast, the film responds almost instantaneously to a bias voltage switch. Therefore, the time constant indeed comes from the IR illumination and not a delay due to an R-C like circuit existing in the entire setup. The dynamic response to the IR source is well described by the exponential $I(t)=I_{light}+D \exp(-t/\tau)$ where $\tau$ is the time constant and D is a scaling constant. From the fit in Figure 3 we find the time constant to be 1.2 seconds. The time constant measured for all our sample ranges from 0.6 to 1.2 seconds and does not show any clear pattern as we increase the loading ratio of MWNT.

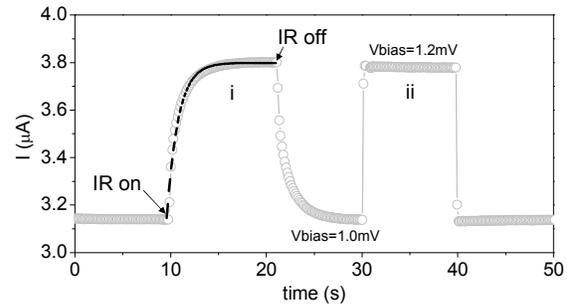

**Figure 3:** Current as a function of time (*t*) for 1.5% film when (i) illuminated at position A (laser is turned on at *t*=10s, kept on for 10s, and then turned off at *t*=20s) and (ii) with a bias voltage spike in dark at *t*=30s, left on at 1.2 mV for 10s and then reduce to 1.0 mV at *t*=40s. This verifies that the time response is due to the IR illumination. The fit is shown as the dashed line which gives a time constant of 1.2 seconds.

All of the results presented here are consistent with the model of Schottky barrier modulation for photocurrent generation [10]. In other words, when the laser light is illuminated at the interface, some energetic electrons overcome the tunnel barrier at the interface and fall into metal electrodes leaving holes in the film. This causes an electron-hole separation at the interface and thereby creates a local electric field. Whereas, when the laser is shined in the middle electron hole pairs are created, however, the charge does not get separated so the overall photovoltage is zero. We rule out both bolometric and molecular photodesportion because the positive and negative photoresponse at two different positions can not be explained by these models. In addition, for bolometric response the time constant is expected to be in the 1-100 ms range [7]. Recently, slow photoresponse in SWNT film has been explained by charge carrier diffusion due to the low mobility of the interconnected SWNT network compared to individual SWNT device. If charge carrier diffusion was responsible in our devices, we would have expected the time constant to decreases as we increase MWNT loading ratio due to increased mobility. Repeated measurements on several sets of samples did not reveal any correlation. Further experimental and theoretical work will be needed to explain the slow time constant in the nanocomposite films.

In conclusion, we present a NIR photoresponse study of large area MWNT/P3HT-b-PS films for different loading ratio of MWNT. We found that upon NIR illumination, the photocurrent is either positive, negative or almost zero based on the position of the illumination with respect to electrodes. The photoresponse is up to 164% when MWNT is embedded into P3HT-b-PS while for pure MWNT film it is only 4%. The time constant of the photoresponse is rather slow, between 0.6-1.2 seconds. Our results are consistent with the model of Schottky barrier modulation for photocurrent generation. This work shows promising novel route for the fabrication of large area low cost infrared photo detectors and position sensitive detectors.

We thank Robert Peale for many useful discussions. This work is partially supported by US National Science Foundation under grants ECCS 0748091 (CAREER) and ECCS 0801924 to SIK, DMR 0746449 (CAREER) to LZ, and DMR 0552295 (CAREER), DMI 0506531 (NIRT) to QH.